# Miniature Annular Permanent Magnet Assembly with Fast Far Field Decay and Majority Collimated Central Field

Garnet Cameron[1,*], Jonathan Cuevas[1], David Shiner[1]

[1] Physics Department, College of Science, University of North Texas, 210 Avenue A, Denton, TX 76203, USA; garnetCameron@my.unt.edu
* Correspondence: garnetCameron@my.unt.edu

**Abstract:** We report on a 9.5 mm (3/8") OD x 6.7 mm (1/4") ID x 9.5 mm (3/8") long NdFeB assembly of 1.6 mm (1/16") thick customized annular magnets boasting quadrupole far field decay making shielding unnecessary; and greater than 50% of ID collimated better than 1/100. The device is applied successfully to a $^{3,4}$He laser spectroscopy experiment.

## 1   Background

An intended application of this magnet assembly is an existing atomic laser spectroscopy experiment that has evolved through at least three versions since 1995 [1] [2] [3] starting from $^4$He only with 51 ppb precision [1]. It is now capable of simultaneous measurements with $^3$He and $^4$He isotopes at 20 ppb [3].

Increasing or decreasing atomic $m_j$ by laser stimulation demands $\Delta m_j = \pm 1$ thereby requiring [X] or [Y] transitions [3]. The laser E-field (polarization) should be perpendicular to the quantization axis (z) defined by the imposed B-field ($B_{magnet}$). As suggested by Kastler [4], laser propagation ($\vec{k}$) parallel to $B_{magnet}$ guarantees this requirement for a circularly polarized laser, as shown in Figure 1. A permanent magnet is a passive and compact magnetic field source versus high current and turns-dense solenoids with associated heating issues.

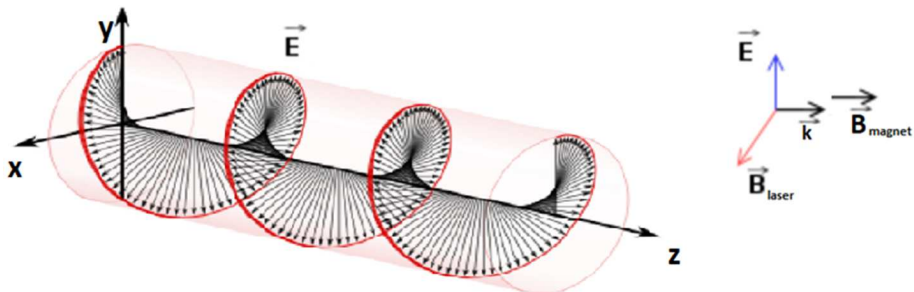

**Figure 1 Atomic Beam Pump Laser Polarization, $\vec{k}$, and B$_{magnet}$ arrangement (Left or $\sigma^+$ shown)**

## 2 Initial Magnet Requirements and Dimensions

Legacy atomic experiment $10^4$ interval count translates to a Poisson noise threshold of 1/100 by $\varepsilon = \sigma / \sqrt{N}$ [5]. This threshold is expected to continue into projected higher signal levels coupled with reduced count interval time length. An ambitious[1] 1/100 directional sensitivity of perpendicular polarization to the atomic beam is sought. Gaussmeter measurements around the neighboring electron gun magnets that use a return flux yoke revealed 5 G perpendicular surrounding field suggesting a 500 G minimum magnetic field strength.

Fringing field close to the probe laser interaction point should be below background i.e. Earth's field. The atomic beam pump laser is approximately 3.5" from the interaction point. The magnet assembly should fringe to background in less than 3". Shielding would impact other needed magnetic fields in the apparatus.

Avoidance of experiment apparatus redesign prefers the magnet assembly to fit in the existing space for linear polarized atomic beam pumping. The critical dimension is 3/8" along the atomic beam line (outer diameter of annular magnet) while providing enough atomic beam pumping transit distance to allow 99% pumping to $m_j$ = +1 or -1 (internal diameter). Figure 2, upper left corner shows a cross-sectional layout.

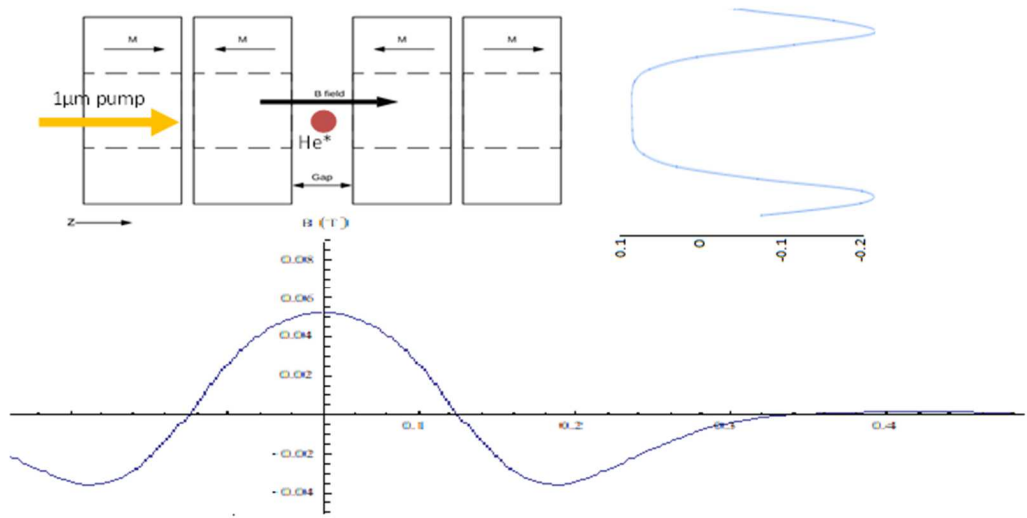

**Figure 2 Magnet Assembly: bottom, axial (z) $B_z$ field; right: mid-plane transverse (x-y) $B_z$ field**

---

[1] Electromagnetic energy is proportional to $B^2$ prompting a more current minimum alignment as 1/10. But, latitude for future order of magnitude precision increases supports 1/30, then 1/100.

Faraday isolators like Gauthier's design [6] offer some inspiration towards the fringe field requirement. As shown in Figure 2, stacked pairs of magnets with opposed magnetization sum to a zero far field. Magnet position difference impact reduces as we move further away. The central field is collimated in Helmholtz fashion as the parallel magnetized pair of magnets straddling the Gap dominates there.

It is important to note that coincident points on <u>axial (z-axis)</u> and <u>transverse (x-axis)</u> B-field profiles within the Gap must have the same value i.e. the center point physically. Axial (z) and transverse (x or y or r) B-field plots are shown in the bottom and right of Figure 2, respectively. The areas beyond the magnets comprise the far field while the Gap is labelled.

## 3  Simple Configuration

Magnetic fields are solved in two ways. A quick axial/longitudinal (z-axis) solution can be determined by Biöt Savart current coil model summation. Alternatively, Finite Element Analysis (FEA) using Maxwell's Equations [7] provides a more complete solution (longitudinal or z-axis, and transverse or x-axis).

$$\vec{H} \equiv -\vec{\nabla}\Phi$$

$$\vec{\nabla} \cdot \vec{B} = 0$$

$$\vec{B} = \mu\left(\vec{H} + \vec{M}\right)$$

Outside magnetic material, M = 0:

$$\vec{\nabla} \cdot \vec{B} = \mu_0\left(\vec{\nabla} \cdot \vec{H} + \vec{\nabla} \cdot \vec{M}\right) \quad \Rightarrow 0 = \mu_0\left(\vec{\nabla} \cdot \left(-\vec{\nabla}\Phi\right) + 0\right) \quad \Rightarrow 0 = \nabla^2\Phi$$

Inside magnetic material of constant magnetization, $\vec{M}$ = constant:

$$\vec{\nabla} \cdot \vec{B} = \mu_0\left(\vec{\nabla} \cdot \vec{H} + \vec{\nabla} \cdot \vec{M}\right) \quad \Rightarrow 0 = \mu_0\left(\vec{\nabla} \cdot \left(-\vec{\nabla}\Phi\right) + 0\right) \quad \Rightarrow 0 = \nabla^2\Phi$$

$\vec{M}$ parallel to the surface or *Dirichlet* boundary condition (BC) is applied to the far bounding surfaces while $\vec{M}$ perpendicular to the surface or *Neuman* BC is employed at symmetry faces[2] involving or very close to the magnet(s) e.g. the floor in Figure 3. The former assumes minimal magnetic domain fan-out in the bulk while the later allows edge effects.

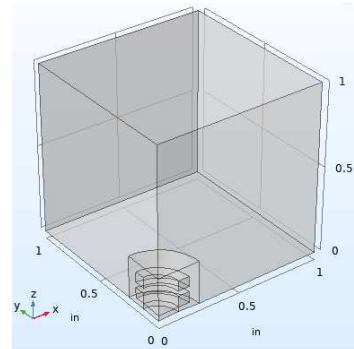

<span style="color:#C0504D">**Figure 3 Reduced FEA Simulation Volume**</span>

Exploration Finite Element Analysis (FEA) simulations shown in Figure 4 examine different configurations of similar annular magnets. Clearly, a pair stabilized the central field. Addition of a repelling pair reduces the far field decay distance.

---

[2] Model space is reduced by exploiting symmetry planes to minimize simulation overhead. The whole solution is generated by duplicating the minimized volume.

Exploration Finite Element Analysis (FEA) simulations shown in Figure 4 examine different configurations of similar annular magnets. Clearly, a parallel magnetized pair stabilized the central field. Addition of a repelling pair (one in each end) reduces the far field decay distance.

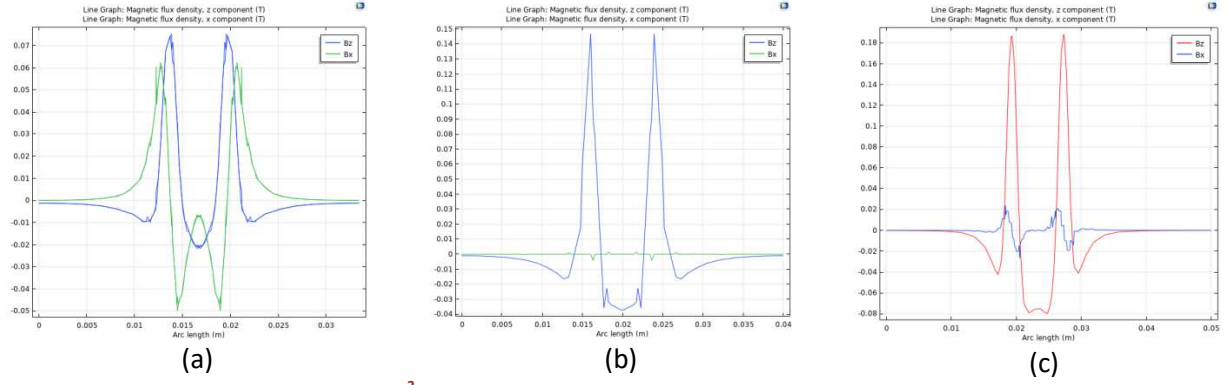

(a)                    (b)                    (c)

**Figure 4 Simulated[3] Transverse $B_z$ and $B_x$ for: (a) Single; (b) Pair; and (c) AR stack.**

## 3.1  B-field Collimation

Magnet dimensions were selected by Biöt Savart current coil approximation simulations[4] [7] exploring axial B-field strength along the magnet assembly z-axis as the Gap size is varied as seen in Figure 6 (a). Each magnet is modeled as a pair of inner and outer coaxial, z-centered, and counter rotating current loops[5]. Summation of infinitesimal magnetization current loops across the magnet bulk yields the current loops model [7].

From the Biöt Savart law:

$$d\vec{B} = kI \frac{\left( d\vec{l} \times \vec{r} \right)}{\left| \vec{r} \right|^3}$$

The well-known analytical solution of axial field for a circular coil is [8]:

$$B_{z,axial} = \frac{\mu_0 I}{2} \frac{r^2}{\left( z^2 + r^2 \right)^{3/2}}$$

Therefore, the assembly axial field $B_z(z)$ is a summation over all the coils/magnets:

$$B_z \left( z \right) = \frac{\mu_0 I}{2} \left\{ \sum_{\substack{i=2,3 \\ j=1,2}} \frac{\left( -1 \right)^j r_j^{\,2}}{\left[ \left( z - z_i \right)^2 + r_j^{\,2} \right]^{3/2}} - \sum_{\substack{i=1,4 \\ j=in,out}} \frac{\left( -1 \right)^j r_j^{\,2}}{\left[ \left( z - z_i \right)^2 + r_j^{\,2} \right]^{3/2}} \right\}$$

---

[3] COMSOL generated. Arbitrary positive and negative B sense is inverted compared the rest of analysis.

[4] Code provided at https://github.com/garnetc/CircularLaserMagnet.

[5] Multiple pairs of loops were simulated and found unnecessary – only a central pair was sufficient.

where, the magnets are numbered left to right, 1 to 4 and are indexed as i. The inner and outer coils are represented by j and are numbered 2 and 1, respectively.

Antiparallel magnetization directions sum depending on Gap size. Axial (z-axis) field strength is also influenced by the ratio of inner radius ($R_i$), outer radius ($R_o$), and thickness of each magnet. In Figure 6 (c), solenoid simulation identifies the optimum thickness and $R_i/R_o$ values. However, thickness and OD selection were constrained by commercially available options: 1/8" or 1/16" and OD ($2R_o$) desired to 3/8" (mentioned earlier), respectively. Conceptually, smaller thickness would be preferable when the annular magnet is modelled as inner and outer solenoids. Thicker magnets relative to outer diameter tend to the infinite coils solution where the axial fields sum vanishes - axial fields of the inner and outer coils are equal and opposite. Jackson's [7][6] solenoid-faces-subtending-angles solution shows this behavior:

$$B_{z,axis}\left(\theta_1, \theta_2\right) = \frac{\mu_0 NI}{2}\left(\cos\theta_1 + \cos\theta_2\right)$$

$$B_{z,axis}\left(\theta_1, \theta_2\right) \xrightarrow[\theta_1 = \theta_2 = 0]{} \mu_0 NI$$

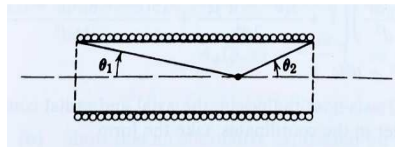

**Figure 5 Thick Magnet Axial Field Behavior**

The current loop model also suggests the OD should be as large as possible i.e. $R_i/R_o$ minimized. Outer loop negative impact on the sum axial ($B_z$) field is reduced. Experiment apparatus 3/8" available space therefore sets the OD.

---

[6] ppg 225

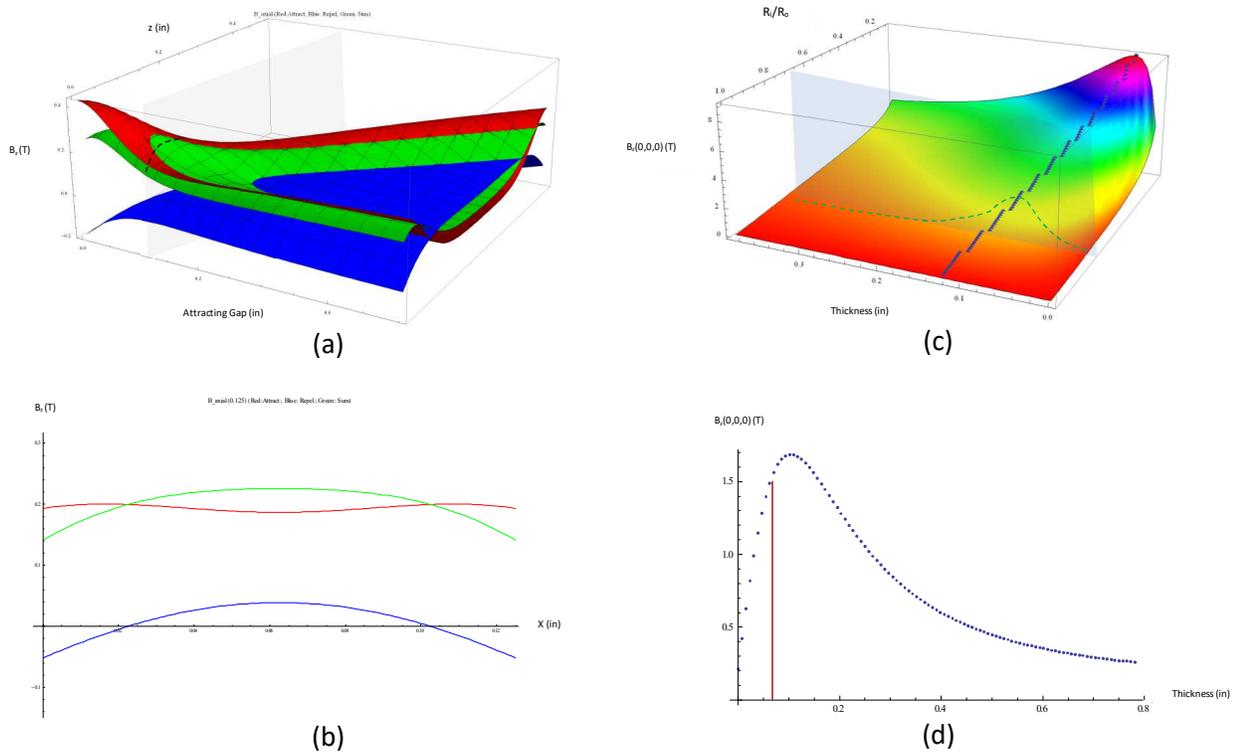

**Figure 6 Simple Biöt Savart Axial (z-axis) Simulation Plots:**
**(a) $B_z(z, Gap)$ – Attracting Pair (Red), Repelling Pair (Blue), Sum (Green), system position slice and sum curve (shaded plane and dotted line);**
**(b) $B_z(z, Gap = 0.125")$[7] – Attracting Pair (Red), Repelling Pair (Blue), Sum (Green);**
**(c) Single Magnet Face Center $B_z(Ri/Ro$, thickness) – maximum values (Black dots), system slice and curve (shaded plane and green broken curve);**
**(d) Single Magnet Face Center $B_z(R_i/R_o = 0.67$, thickness) – system position (red line)**

3/8" OD constraint and commercial unavailability of 3/8" x 1/4" x 1/16" forced customization of 3/8" x 1/8" x 1/16" via Dremel-lathe working to widen the 1/8" ID to 1/4". Commercial annular magnets are sintered powders encased in a thin metallic jacket; in this case Ni-Cu-Ni. Usual cutting tools result in fracturing and disintegration. Grinding is the only viable option. Colloidal graphite was used to stabilize the exposed surfaces[8]. An Ultra-High Vacuum (UHV) chamber compatible mount was manufactured retaining the 1/16" gap, illustrated in Figure 7. 80 mil and 180 mil vertical gap versions allowed vacuum experiment and bench probing, respectively. Profile results are shown in Figure 8 revealing a 0.1" to 0.15" collimated B-field diameter.

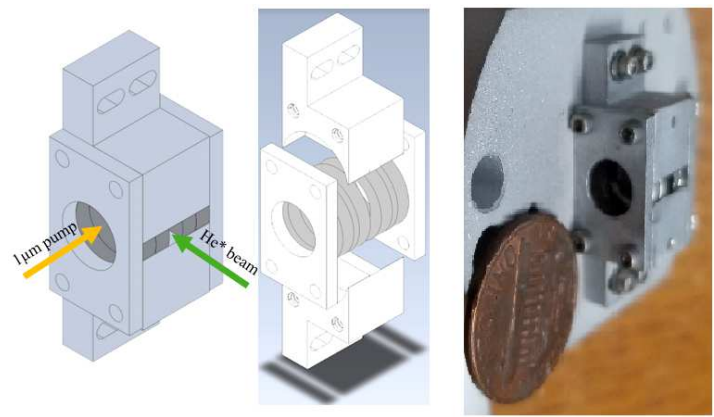

**Figure 7 UHV Atomic Beam Pump Magnet Mount**

---

[7] 0.125" Gap size due to coils modeled at the center point of 1/16" thick magnet i.e. actual Gap (1/16") + 2 * coil offset (1/32")
[8] Colloidal graphite is also useful to minimize stray pump laser reflections.

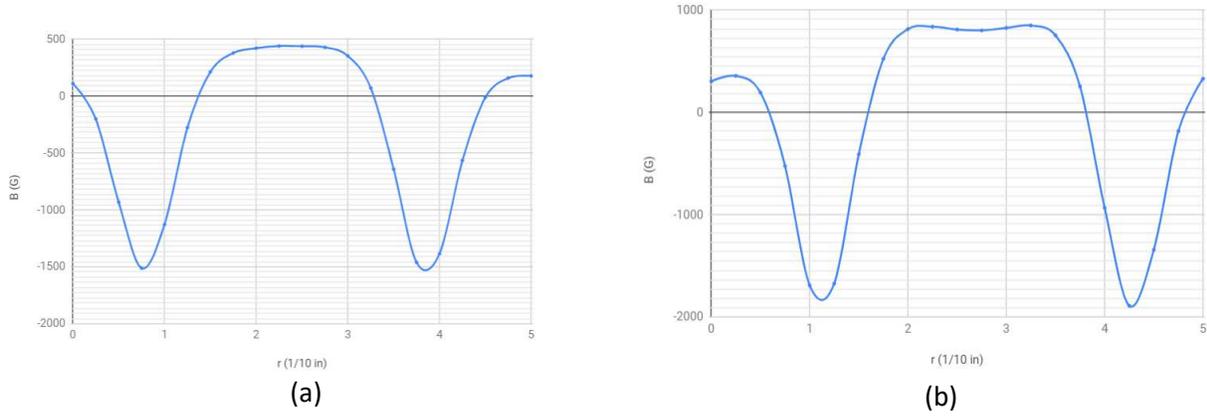

(a)

(b)

**Figure 8 Atomic Beam Pump Laser, 3/8" x 1/4" x 1/16" Magnet Transverse B-field Profiles: (a) Single; (b) Assembly**

1/4" ID was selected based on minimum atom travel distance, d, dictated by the number of atomic beam cycles, n, to allow 1/100 depletion of $2^3S_1$:$m_j$=0 level. The pump/applied laser will be broad bandwidth covering all possible magnetic transitions ($\Delta m_j = 0$). A simplistic estimate is achieved from the single excitation remaining ratio[9], r, excitation life time, $\tau$, number of excitations/cycles, n, and atom velocity, $v_{rms}$.

$$N = N_0 r^n \Rightarrow n = \log\left(\frac{N}{N_0}\right)\Big/\log r$$

$$d = v_{rms}\left(n\tau\right) = v_{rms}\,\tau \log\left(\frac{N}{N_0}\right)\Big/\log r$$

r can be assessed[10] by examination of excitation and relaxations paths assuming similar transition probabilities and branching rations. Traditional transitions [9] are shown in Figure 9 with ten excitation paths (a) and six relaxation paths (c) to $2^3S_1$:$m_j$=0 yielding r as 0.6.

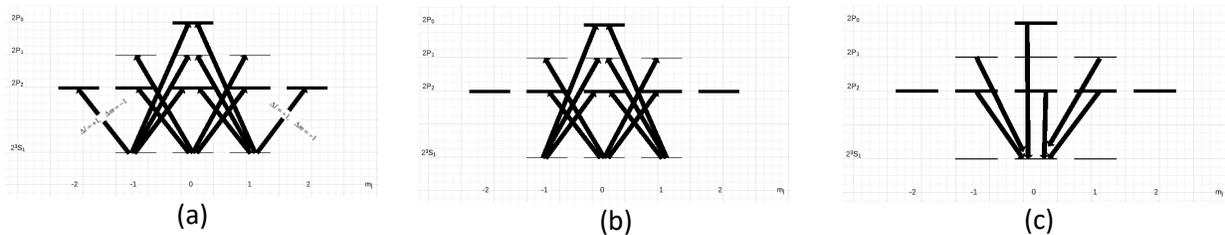

(a)

(b)

(c)

**Figure 9 Linearly Polarized 1 μm Laser Excitation and Relaxation Branching of $^4$He\*: (a) All Possible Excitations Paths; (b) Excitation to $2^3S_1$:$m_j$=0 Relaxation Levels; (c) $2^3S_1$:$m_j$=0 Relaxation Paths[11].**

---

[9] Portion of atoms still in $2^3S_1$:$m_j$=0 state.

[10] Using a simplified model that assumes all transition matrix elements are similar.

[11] $2^3P_1$:$m_j$=0 → $2^3S_1$:$m_j$=0 transition is dark due to [Z] ($\Delta l = \pm 1$, $\Delta m = 0$) and [X]/[Y] ($\Delta l = \pm 1$, $\Delta m = \pm 1$) selection rules.

$$v_{rms} \cong 1500 \, m/s$$

$$\tau = 100 \, ns$$

$$N/N_0 = 0.01$$

$$r = 0.6$$

$$d = 1500 \times 1E-7 \times \log(0.01)/\log(0.6) = 1.352 \, mm = 0.05 \, in$$

Minimum 0.05 inch He* beam travel distance requirement fits well with 0.15 in diameter of collimated[12] imposed/quantization B-field.

### 3.1.1 Atomic Beam B-Field Collimation Precision

It is important to note that differences between Biö Savart simulation (Figure 6 (b)) and the experimental (Figure 8 (b)) field profile shapes stems from axial versus transverse independent variable - z versus x or y[13], respectively.

The rectangular atomic beam flows in the x-direction. It is 2 mils wide in the longitudinal direction (z) and 80 mils (2 mm) tall (y). Longitudinal field variation has small effect across the 2 mil range. But, transverse field variation has significant impact across the possible 80 mills expanse depending on the pump laser beam diameter. The empirically measured dip shown in Figure 8 (b) is 50 Gauss compared to 800 Gauss at the shoulders (~6%) over the 150 mil collimated B-field widow can be of concern.

## 3.2 3-D Simulations

3-D simulations model more completely the physical situation (despite z-axis rotational symmetry) therefore providing more accurate values, and both axial (z-axis) and transverse (x-axis) values. EMS[14] and COMSOL (along with the Biö Savart summations and experimental readings) were used to produce the longitudinal (z-axis) and transverse plots (x-axis) shown in Figure 10.

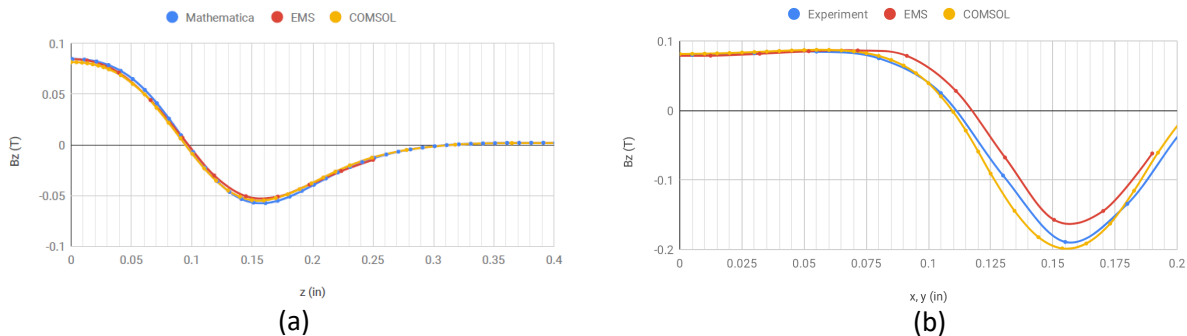

(a)                    (b)

**Figure 10 B$_z$ Combined Plots – Origin at symmetry axis: (a) Longitudinal (z-axis)[15]; (b) Transverse (x-axis)**

---

[12] Constant B-field measured along the transverse (x) direction implies collimated B-field in the axial (z) direction.

[13] Z-axis rotation symmetry.

[14] Solid Works add-in by EM Works.

[15] Google Sheets resolution does not display ~60 G dip well at the origin.

The simple axial/longitudinal Biö Savart Mathematica model compares favorably with EMS and COMSOL simulations. Similarly, EMS and COMSOL fit well with transverse experimental measures. The 50 Gauss central dip is apparent in the transverse plot.

## 3.3   Far Field

Minimizing impact (below noise threshold of ~1/100) at the 3" distant interaction region is another major requirement for this magnet. ~800 G central B-field translates to ~8 G fringe field threshold. Significant threshold fringe field distance differences are noted in for small spacing changes in Attracting Pair (AP) and Attracting-Repulsion (AR) boundaries as shown in Table 1 and Table 2. Greater attention is paid to the transverse far field behavior because the interaction volume is largely transverse to the magnet assembly. However, axial or z-axis expected behavior is included for completeness.

**Table 1 Attracting Pair (AP) Separation Varied with Touching Attracting-Repelling (AR) Boundary**

| AP (in) | Axial (z-axis) Simulation | Transverse (x-axis) Empirical Measurement[16] | Empirical 8 G, Knee |
|---------|---------------------------|-----------------------------------------------|---------------------|
| 1/16 |  Zoomed Far Field B_axial (0.0625, z) (Red :Attract ; Blue : Repel; Green : Sum |  Far Field 3/8 x 1/4 x 1/16 Assembly (Center 20") | ~1", ~2" |
| 2/16 |  Zoomed Far Field B_axial (0.125, z) |  Far Field 3/8 x 1/4 x 1/16 Assembly with 1/16 AP Spacer (Center 20") | ~0.25", ~0.5" |

---

[16] Earth's field remains in the data and flips sign due to probe orientation.

**Table 2 Attracting-Repelling (AR) Boundary varied with 1/16" Attracting Pair (AP) Separation**

| AR (in) | Simulation | Empirical | Empirical 8 G, Knee |
|---------|-----------|-----------|---------------------|
| 1/32 | 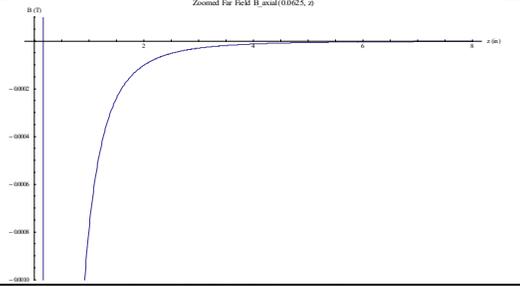 | 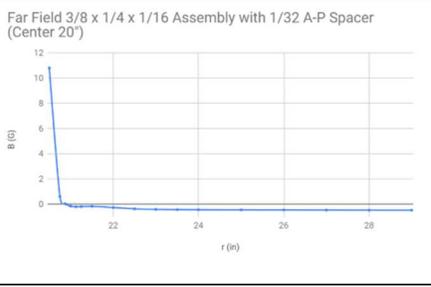 | ~0.25", ~0.5" |
| 1/16 | 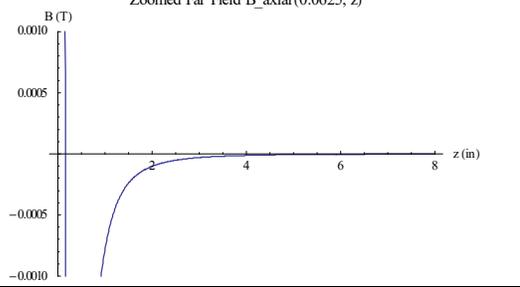 | 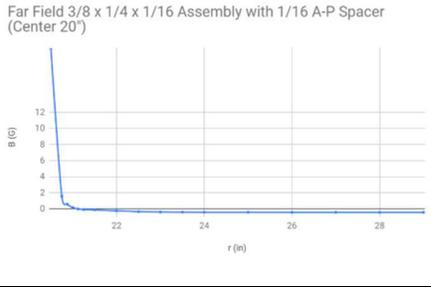 | ~0.5", ~0.75" |

1/16" AP gap with no AR space (touching) meets the 3" far field requirement with better performance in all other configurations presented.

Transverse (x) asymptotic behavior was explored using the 1/16" AP gap and touching AR scenario – the slowest decay configuration. Figure 11 expands the inflection point seen in Table 1 (first row, second column) and compares the $B_z$-field with $1/r^4$ decay (red broken line). Field dissipation is approximately quadrupole.

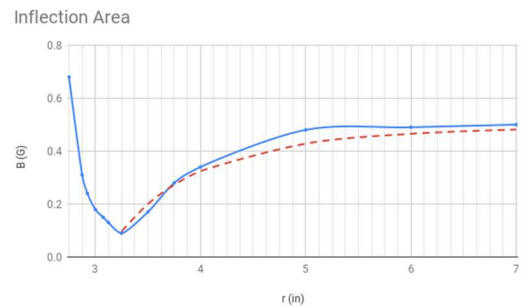

**Figure 11 Far Field Asymptotic Behavior and $1/r^4$ (red line)**

# 4    Application Optimization

Application of the simple[17] stack configuration to intended atomic laser stimulation focuses the B-field-collimation requirement to 90% $B_z$ in a transverse (x-y plane), 2 mm diameter, co-axial circular area centered in the Gap. Perpendicular-polarized laser stimulation can only be 10% impure and the transverse (x-y plane) extent is the most restrictive since the atomic beam in only 2 mils (0.05 mm) in the magnet assembly longitudinal direction; but, 80 mils (2 mm) in the traverse direction. <u>$B_{x,y}$ needs to be less than 8 G with respect to 800 G $B_z$.</u>

Revisiting Figure 6 (a) and (b) with an eye to the repelling pair impact on the Gap B-field shows an antagonistic relationship i.e. too close depresses the Gap field. Introducing spacing between the

---

[17] No AR spacing.

repelling and attracting magnet faces or AR-space, is indicated. Using standard material available, 1/32"
thick SS-316 washers were implemented as spacers on each AR boundary. Empirical data confirms this
insight shown in Figure 12.

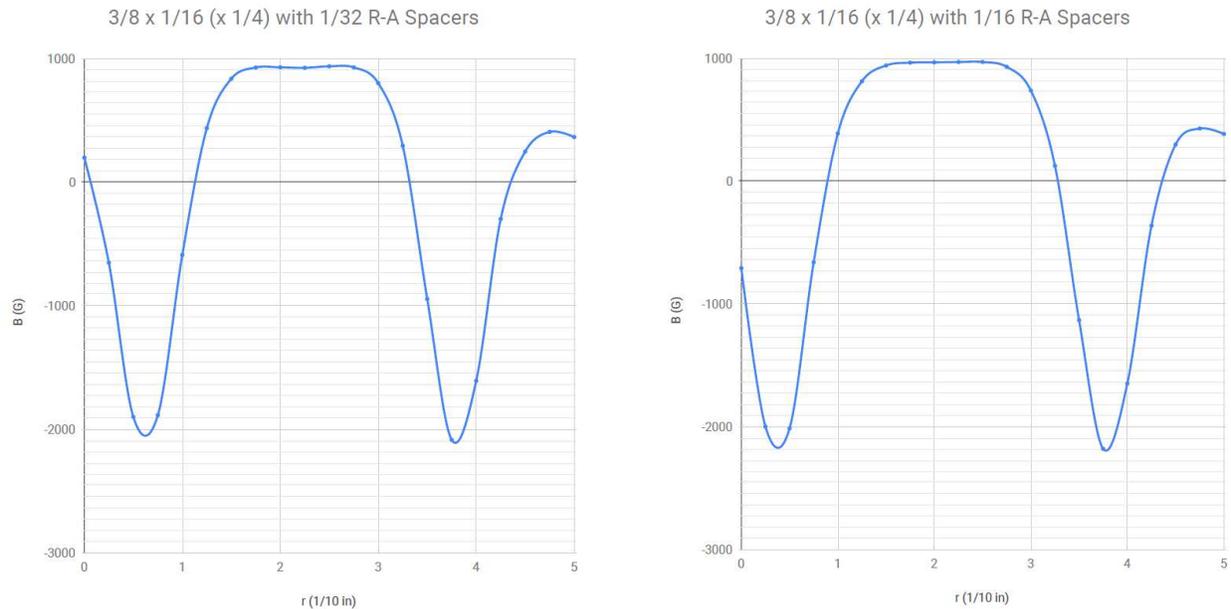

<div align="center">

**Figure 12 AR Spaced Traverse B$_z$: (a) 1/32"; (b) 1/16".**

</div>

1/32" AR spacing shows a 1/450 dip (vs ~1/10 with no AR space) while 1/16" improves slightly to 1/500.
But, the "constant" B$_z$ field area narrows in the latter from 0.1" to 0.75"[18].

Further, AR spacing influences far field decay distances. Table 1 and Table 2 show 8 G decay distances
for 0, 1/32", and 1/16" spacing to be ~2", ~0.5", and ~0.75", respectively. 1/32" AR spacing is reinforced
as an optimal choice.

Initial experiment magnet use showed 40% pumping efficiency versus a 99.99%[19] expected value.
Further study and collimation improvement were inferred. Close simulations of B$_x$ behavior for both no
AR spacing and 1/32" AR spacing were conducted for varying transverse (x-y plane) offsets to center (z =
0), as shown in

Figure 13.

---

[18] 1/16" AR spacing is an area for further study – cost benefit/tradeoff on smaller B$_z$ dip vs narrower usable region.
[19] Leading term in transition is B$^2$.

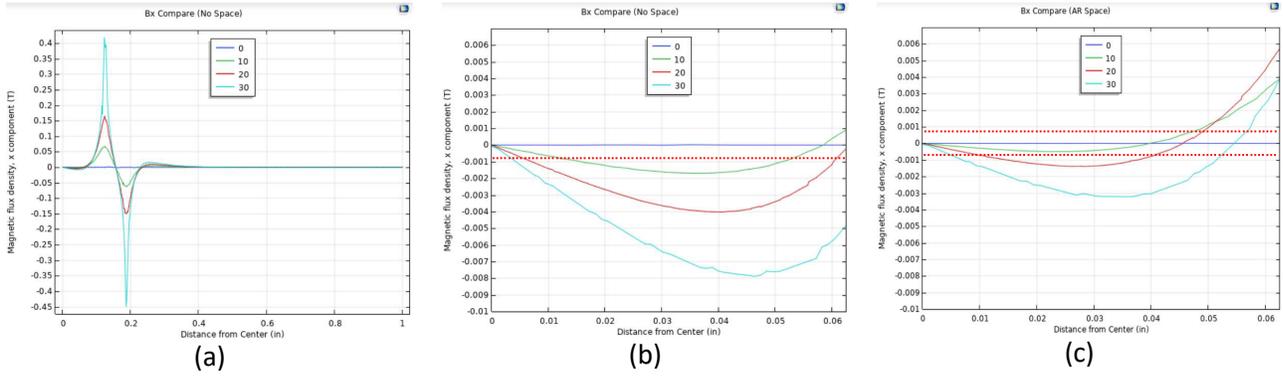

(a)    (b)    (c)

**Figure 13 B_x(x or y) at Varying Longitudinal Offset (in mils): (a) Wide; (b) No AR space (c) 1/32" AR space**

The broken red line identifies the 8 G transition efficiency threshold. Clearly, the AR-space configuration is acceptable at 10 mils while the no-AR-space setup does not. The former is selected with the proviso of ±10 mils magnet transverse (x-y plane) offset from center (z = 0) alignment with the atomic beam and 90 mils/2 mm diameter stimulation laser beam.

# 5    Performance

The figure of merit for He* preparation is the pumping ratio – portion of $2^3S_{1,\ m_j = 0}$ removed. This requires singlet quenching to isolate triplets. Photon background levels can offset the zero signal level being pursued. Figure 14 shows the raw pumping ratio with photon background. Two features are apparent: pumping ratio is scaling with power i.e. not saturated; and the best ratio (83%) is actually higher – ~96%[20] using measured 12% photon background[21]. The current core-pumped preparation laser's 980 nm Bookham diode laser pump is rated for 1000 mA (maximum exceeded by ~50% at ~300 mW output) hindering higher power application. A higher power preparation laser is needed.

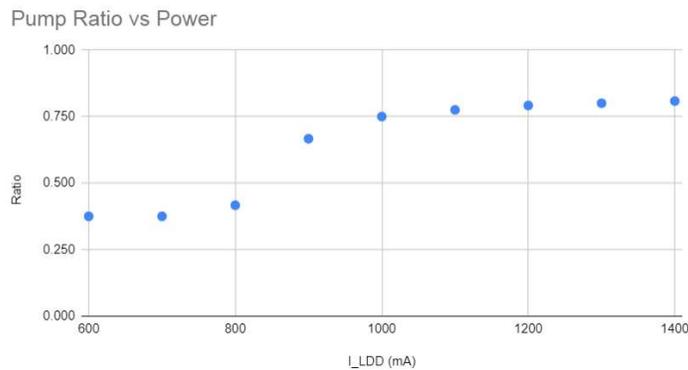

**Figure 14 Preparation Laser Raw Pumping Ratio (1µm, σ-/right) vs Power (LDD current)**

---

[20] Zero/near-zero B-field transit presents a pumping obliteration/reduction challenge as Zeeman energy level separations disappear/narrow.

[21] Preparation and Probe laser combination method demonstrates 12% background.

# 6 Conclusions and Further Work

This miniature magnet has been shown the present an 800 G, well collimated axial B-field (z-axis) to better than 1/100 as verified by simulation and atomic experiment. The fringe field falls asymptotically as quadrupole ($1/r^4$) reaching background in less than 3" (7.5 mm) known by simulation and experiment also. Any space constrained application needing a non-shielded, localized, and collimated magnetic field source can use this device or like design.

Strong improvements possible from this effort are:

1. ***Higher power laser beyond 300 mW*** can improve the unsaturated 96% measurement channel clearing to an impeccable 99.99%;

2. ***2/16" or greater AP gap*** may further improve the far field decay rate. Concerns about axial (z-axis) B-field size summarily reduced this option's importance in this study; and

3. ***1/16" or greater AR spacing*** holds the potential of reduced $B_x$ affording better collimation purity.

Items 1 and 3 hold great promise beyond current performance.